\documentclass{aastex631}

\usepackage{soul}
\submitjournal{PSJ}

\shorttitle{Physical Characterization of Metal-rich NEAs}
\shortauthors{Sanchez et al. 2021}
\graphicspath{{./}{figures/}}

\begin{document}

\title{Physical Characterization of Metal-rich Near-Earth Asteroids 6178 (1986 DA) and 2016 ED85}

\correspondingauthor{Juan A. Sanchez}
\email{jsanchez@psi.edu}

\author{Juan A. Sanchez}
\altaffiliation{Visiting Astronomer at the Infrared Telescope Facility, which is operated by the University of Hawaii under Cooperative Agreement no. NNX-08AE38A with 
the National Aeronautics and Space Administration, Science Mission Directorate, Planetary Astronomy Program.}
\affiliation{Planetary Science Institute, 1700 East Fort Lowell Road, Tucson, AZ 85719, USA}

\author{Vishnu Reddy}
\altaffiliation{Visiting Astronomer at the Infrared Telescope Facility, which is operated by the University of Hawaii under Cooperative Agreement no. NNX-08AE38A with 
the National Aeronautics and Space Administration, Science Mission Directorate, Planetary Astronomy Program.}
\affiliation{Lunar and Planetary Laboratory, University of Arizona, 1629 E University Blvd, Tucson, AZ 85721-0092}

\author{William F. Bottke}
\affiliation{Southwest Research Institute, Suite 300 1050 Walnut Street, Boulder, CO 80301, USA}

\author{Adam Battle}
\affiliation{Lunar and Planetary Laboratory, University of Arizona, 1629 E University Blvd, Tucson, AZ 85721-0092}

\author{Benjamin Sharkey}
\affiliation{Lunar and Planetary Laboratory, University of Arizona, 1629 E University Blvd, Tucson, AZ 85721-0092}

\author{Theodore Kareta}
\affiliation{Lunar and Planetary Laboratory, University of Arizona, 1629 E University Blvd, Tucson, AZ 85721-0092}

\author{Neil Pearson}
\affiliation{Planetary Science Institute, 1700 East Fort Lowell Road, Tucson, AZ 85719, USA}

\author{David C. Cantillo}
\affiliation{Lunar and Planetary Laboratory, University of Arizona, 1629 E University Blvd, Tucson, AZ 85721-0092}








\begin{abstract}

Metal-rich near-Earth asteroids (NEAs) represent a small fraction of the NEA population that is mostly dominated by S- and C-type asteroids. Because of this, their identification and study provide us with a unique opportunity to 
learn more about the formation and evolution of this particular type of bodies, as well as their relationship with meteorites found on Earth. We present near-infrared (NIR) spectroscopic data of NEAs 
6178 (1986 DA) and 2016 ED85. We found that the spectral characteristics of these objects are consistent with those of metal-rich asteroids, showing red slopes, convex shapes, and a weak pyroxene absorption band at $\sim$0.93 $\mu$m. The compositional analysis showed that they have a pyroxene chemistry of Fs$_{40.6\pm3.3}$Wo$_{8.9\pm1.1}$ and a mineral abundance of $\sim$15\% pyroxene and 85\% metal. We determined that these objects were likely transported to the near-Earth space via the 5:2 mean motion resonance with 
Jupiter. Asteroid spectra were compared with the spectra of mesosiderites and bencubbinites. Differences in the NIR 
spectra and pyroxene chemistry suggest that bencubbinites are not good meteorite analogs. Mesosiderites were found to have a similar 
pyroxene chemistry and produced a good spectral match when metal was added to the silicate component. We estimated that the amounts of Fe, Ni, Co, and the platinum group metals present in 1986 DA could exceed the reserves worldwide.

\end{abstract}




\section{Introduction} \label{sec:intro}

The study of near-Earth asteroids (NEAs) offers the opportunity to look into more detail at the physical properties and composition of their counterparts in the main asteroid belt. NEAs also represent a direct link between 
meteorites found on Earth and their parent bodies in the solar system. Thus, an important part in tracing the origin of these meteorites involves determining the composition and source region of NEAs.

Here, we present near-infrared (NIR) spectroscopic data of NEAs 6178 (1986 DA) and 2016 ED85. This work was motivated by a close flyby of 1986 DA in April 2019. To date, this is the only NEA that has been 
confirmed to be a metal-rich body from radar observations \citep{1991plas.rept..174O}; however, no detailed compositional analysis using spectroscopic data has been performed. During the course of this research, 
we also had the opportunity to observe 2016 ED85 in September 2020 as part of an ongoing survey of NEAs. This object was chosen for observation not only because of its close approach to Earth but also because it 
came from the same region in the outer belt as 1986 DA. The data reduction later revealed that 2016 ED85 had almost identical spectral characteristics to 1986 DA and other known metal-rich asteroids; hence, the asteroid was 
included in our original study. 

Metal-rich asteroids are thought to represent the exposed cores of differentiated asteroids whose crusts and mantles were stripped away following a catastrophic disruption \citep[e.g.,][]{1989aste.conf..921B}. A more recent theory 
suggests that some of these objects, in particular (16) Psyche, might still preserve a rocky mantle, and that the metal present on the surface could be the result of ferrovolcanic eruptions that covered the rocky material with liquid 
metal \citep{2020NatAs...4...41J}. Some of the largest known metal-rich asteroids are located in the middle and outer part of the asteroid belt between $\sim$ 2.65 and 3.0 au. These asteroids have high radar albedos ($\hat{\sigma}_{OC}$), with measured values ranging from $\sim$ 0.22 to 0.60 \citep[e.g.,][]{2007Icar..186..126M, 2010Icar..208..221S, 2015Icar..245...38S} and geometric albedos ($P_{V}$) ranging from $\sim$0.10 to 0.30 \citep[e.g.,][]{2004AJ....128.3070C, 2010Icar..210..674O, 2011M&PS...46.1910H}. They are normally classified as M-types in the Tholen taxonomy \citep{1984PhDT.........3T}, or as Xk- and Xe-types in the Bus-DeMeo taxonomy \citep{2009Icar..202..160D}. In the NIR, their spectra are characterized by having red slopes, convex shapes, and in some cases weak absorption bands at $\sim$ 0.9 and 1.9 $\mu$m attributed to the presence of pyroxene \citep{2010Icar..210..674O, 2011M&PS...46.1910H, 2014Icar..238...37N}. Evidence for a 3 $\mu$m hydration absorption band has also been found on some of these objects \citep[e.g.,][]{2000Icar..145..351R, 2015Icar..252..186L, 2017AJ....153...31T}. Based on their spectral characteristics, metal-rich asteroids have been considered as the possible parent bodies of iron meteorites, enstatite chondrites, stony-irons, and metal-rich carbonaceous chondrites \citep[e.g.,][]{1979aste.book..688G, 1989aste.conf..921B, 1991plas.rept..174O, 2005Icar..175..141H, 2010Icar..208..221S, 2011M&PS...46.1910H}.

In this work we carry out a comprehensive analysis of the NIR spectra of 1986 DA and 2016 ED85 in order to constrain their surface composition. We also use the orbital parameters of 
these objects to determine their most likely source region. In addition, the spectra of the NEAs are compared with the spectra of metal-rich asteroids in the main belt to identify their possible parent body. 
Laboratory spectra of meteorite samples have also been acquired in order to investigate the relationship 
between these NEAs and stony-iron meteorites and metal-rich carbonaceous chondrites. Finally, we estimate the amounts of metals that could be present in 1986 DA and how much they could be worth. 

\section{Size, Radar Albedo, and Surface Bulk Density}

\cite{1991plas.rept..174O} carried out radar observations of 1986 DA with the Arecibo Observatory's 2380 MHz radar. They estimated a radar cross-section ($\sigma_{OC}$) of 2.40$\pm$0.36 km$^{2}$ for this asteroid. The 
radar albedo of an asteroid can be determined if its $\sigma_{OC}$ and diameter ($D$) are known:  

\begin{equation}
\hat{\sigma}_{OC}=\frac{4\sigma_{OC}}{\pi D^{2}}
\end{equation}

Using a diameter of 2.3 km from \cite{1987AJ.....93..738T}, \cite{1991plas.rept..174O} calculated a $\hat{\sigma}_{OC}$ = 0.58 for 1986 DA. This value, when compared to other metal-rich asteroids, is at the highest end of radar 
albedos measured for these objects \citep[e.g.,][]{2010Icar..208..221S, 2015Icar..245...38S}. 

The diameter of 1986 DA has been updated since the work of \cite{1987AJ.....93..738T}. \cite{2011ApJ...743..156M} calculated a diameter of 3.199$\pm$0.381 km for 1986 DA from data obtained with NEOWISE. This value was later 
revised by \cite{2014ApJ...785L...4H}, who obtained a diameter of 2.8$\pm$0.42 km. In the present work we adopt this diameter for our analysis. Thus, inserting this value in Equation (1) gives us $\hat{\sigma}_{OC}$ = 0.39$\pm$0.13 for 
1986 DA. This is lower than the value calculated by \cite{1991plas.rept..174O}, but it is still high enough for this object to be considered an asteroid dominated by metal \citep{2015Icar..245...38S}.

\cite{1991plas.rept..174O} also reported a broad range of $\sigma_{OC}$ values for 1986 DA, ranging from 1.1 to 4.8 km$^{2}$. This range is equivalent to a radar albedo variation across the surface of 0.18-0.78 (assuming $D$ = 2.8 km). \cite{1991plas.rept..174O} attributed the variation in $\sigma_{OC}$ to an extremely irregular surface at a scale of 10-100 m. They concluded that 1986 DA had a nonconvex irregular shape, possibly bifurcated.

\cite{1985Sci...229..442O} found a relationship between the Fresnel radar reflectivity and the bulk density ($\rho$) of particulate mixtures of rock and metal. \cite{2010Icar..208..221S} expressed this relationship as a function of 
the radar albedo (for $\hat{\sigma}_{OC}$$>$0.07):

\begin{equation}
\rho = 6.944\hat{\sigma}_{OC}+1.083
\end{equation}

\begin{figure*}[!ht]
\begin{center}
\includegraphics[height=9cm]{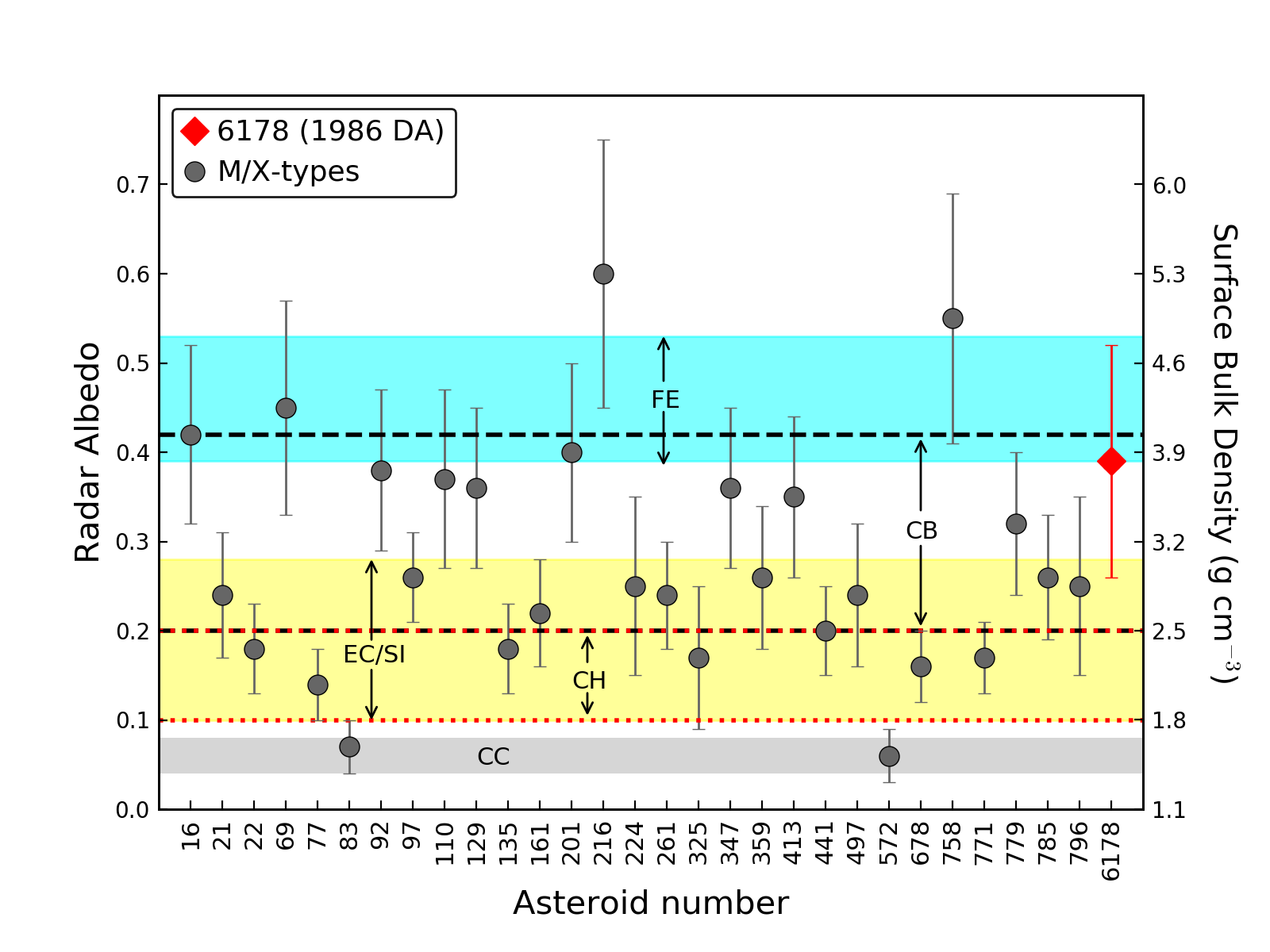}

\caption{\label{f:Figure1} {\small Radar albedo and near-surface bulk density for 1986 DA and all main belt M/X-type asteroids observed by radar. Also shown, regions corresponding to carbonaceous chondrites (CC), high 
metal carbonaceous chondrites (CH), enstatite chondrites (EC), stony-irons (SI), bencubbinites (CB), and iron dominated meteorites (FE). Error bars for the M/X-types correspond to the uncertainties in radar albedo reported by 
\cite{2015Icar..245...38S}. Variations in radar albedo with rotation phase are not shown. Figure adapted from \cite{2010Icar..208..221S, 2015Icar..245...38S}.}}

\end{center}
\end{figure*}

We used this relationship and found that 1986 DA has a surface bulk density of 3.79 g cm$^{-3}$, with a total range of 2.33-6.5 g cm$^{-3}$. According to \cite{2010Icar..208..221S}, surface irregularities on metal-rich asteroids are more likely to cause large variations in radar albedo than on rocky bodies.  Figure \ref{f:Figure1} shows the radar albedo and 
near-surface bulk density for 1986 DA and all main belt M/X-type asteroids observed by radar \citep{2010Icar..208..221S, 2015Icar..245...38S}. The range of possible surface bulk densities for 1986 DA is consistent with most of the values measured for the proposed meteorite analogs for metal-rich bodies. 

Contrary to 1986 DA, there are no radar data available for 2016 ED85. For this reason, in the present study this object will be considered as a candidate metal-rich body. The only parameter available 
for this object is its absolute magnitude, which from the JPL Small-Body Database has a value $H$=17.64. If the absolute magnitude and geometric albedo of an asteroid are known, the diameter can 
be estimated using the following relationship \citep{2007Icar..190..250P}:

\begin{equation}
D(km) = [1329/(P_{V})^{1/2}]\times10^{-H/5}
\end{equation}

Thus, if this object is confirmed to be a metal-rich asteroid, given the range of geometric albedos measured for these bodies, its diameter could be in the range of $
\sim$0.72-1.25 km. Since the radar albedo of 2016 ED85 has never been measured, we cannot estimate its surface bulk density. Future radar observations will be required 
to determine whether the surface properties of this asteroid are similar to those of 1986 DA.

\section{Near-Infrared Spectroscopic Observations}

The two NEAs were observed with the SpeX instrument \citep{2003PASP..115..362R} on NASA's Infrared Telescope Facility (IRTF). NIR spectra (0.7-2.5 $\mu$m) of 1986 DA and 2016 ED85 were obtained in low-resolution 
(R$\sim$150) prism mode with a 0.8” slit width on April 09, 2019 and September 22, 2020 UTC, respectively. During the observations, the slit was oriented along the parallactic angle in order to minimize the effects of 
differential atmospheric refraction. Spectra were obtained in two different slit positions (A-B) following the sequence ABBA. In order to correct the telluric bands from the asteroid spectra, a G-type local extinction star was 
observed before and after the asteroid. NIR spectra of a solar analog were also obtained to correct for possible spectral slope variations. Observational circumstances for the asteroids are presented in Table 1. All spectra were 
reduced using the IDL-based software Spextool \citep{2004PASP..116..362C}. For a detailed description of the data reduction process, see \cite{2013Icar..225..131S}.

\begin{table}[h!]
\caption{\label{t:Table1} {\small Observational circumstances. The columns in this table are: object number and designation, date, phase angle ($\alpha$), V-magnitude, heliocentric distance (r), airmass and solar analog 
used.}}

\begin{tabular}{ccccccc}
\tableline

Object&Date (UT)&$\alpha$ $(^{\circ})$&mag. (V)&r (au)&Airmass&Solar Analog \\  \hline
6178 (1986 DA)&09-April-2019&55&17.4&1.23&1.29&SAO 93936 \\
2016 ED85&22-Sept-2020&43&16.6&1.17&1.07&SAO 93936 \\

\tableline
\end{tabular}
\end{table}

\section{Composition}

\begin{figure*}[!ht]
\begin{center}
\includegraphics[height=9cm]{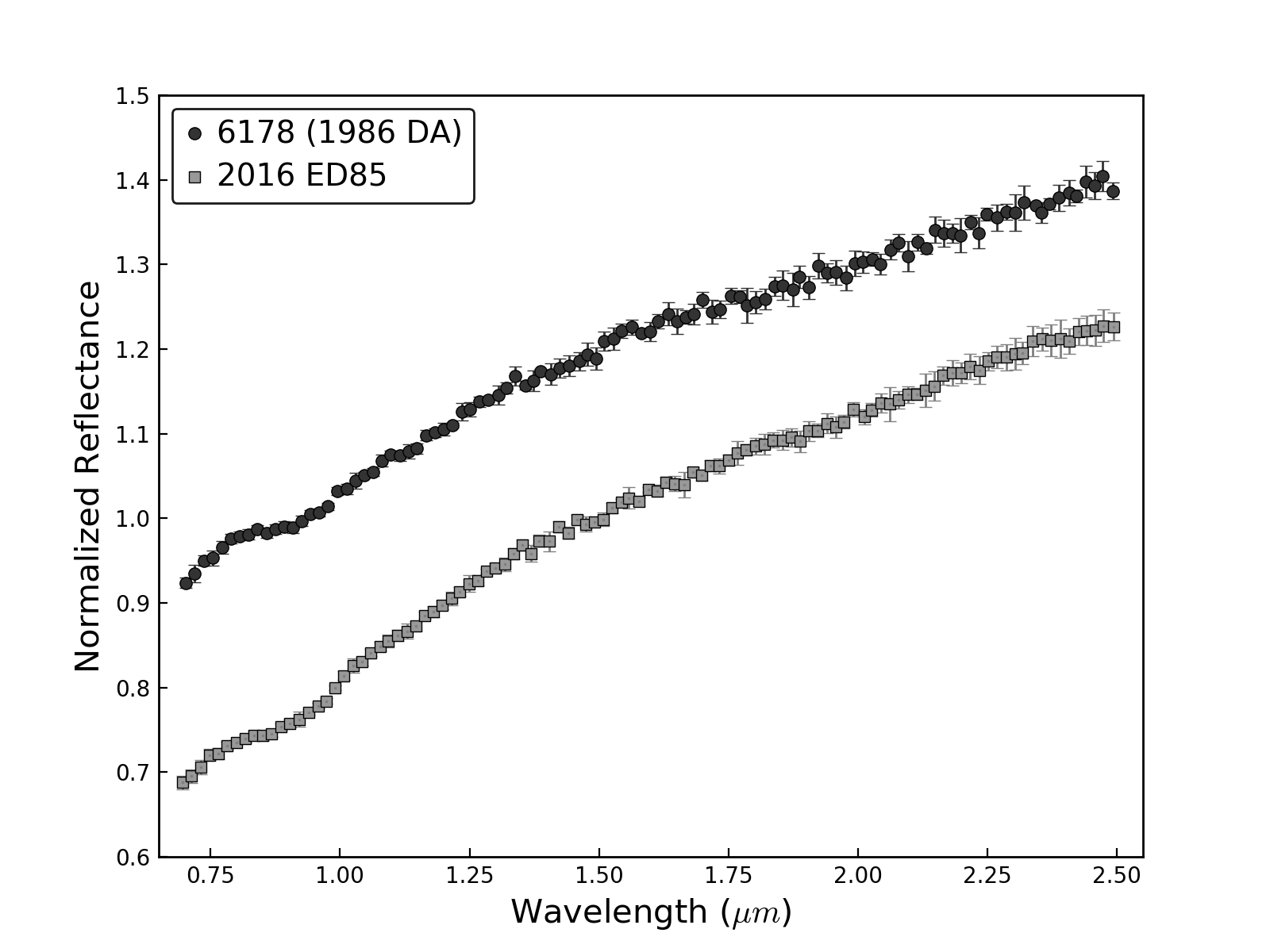}

\caption{\label{f:Figure2} {\small Spectra of 6178 (1986 DA) and 2016 ED85 obtained with the SpeX instrument on NASA’s IRTF.  The spectra have been binned and offset for clarity.}}

\end{center}
\end{figure*}

Figure \ref{f:Figure2} shows the NIR spectra of 1986 DA and 2016 ED85. Both spectra exhibit a red slope and a weak pyroxene absorption band at $\sim$0.9 $\mu$m. The spectra of these two asteroids are very similar, with 
2016 ED85 showing a redder slope than 1986 DA. Spectral band parameters, including the band center and band depth of the 0.9 $\mu$m pyroxene band, were measured from the spectra of 1986 DA and 2016 ED85 using a 
Python code described in \cite{2020AJ....159..146S}. The Band I center was measured after dividing out the linear continuum and corresponds to the position of the minimum reflectance value obtained by fitting a polynomial over the 
bottom of the absorption band. The Band I depth was measured from the continuum to the band center, and it is given as a percentage depth. We obtained a Band I center of 0.93$\pm$0.01 $\mu$m for both objects. The Band I depth of 
1986 DA and 2016 ED85 was found to be 4.1$\pm$0.2\% and 4.3$\pm$0.2\%, respectively.

As mentioned before, 2016 ED85 can only be considered a candidate metal-rich body based on its spectral similarities to 1986 DA and other metal-rich asteroids. In the following analysis we investigate the possibility that the spectral 
characteristics of this object are dominated by the presence of metal, i.e., it will be treated in the same way as 1986 DA, but keeping in mind that these results need to be confirmed by radar observations of this asteroid.

In order to determine the pyroxene chemistry of the NEAs, we employed the equations of \cite{2009M&PS...44.1331B}; these equations were derived from the analysis of howardite, eucrite and diogenite (HED) meteorites and make use of the 
band centers to calculate the molar content of ferrosilite (Fs) and wollastonite (Wo). Both band centers (Band I and Band II) can be used for this calculation. Before determining the pyroxene chemistry, a temperature correction derived 
by \cite{2012Icar..217..153R} was applied to the Band I center in order to account for the differences between the surface temperature of the asteroid and the room temperature at which the equations were derived. A similar 
approach was used by \cite{2017AJ....153...29S} to determine the composition of asteroid (16) Psyche, the largest known M-type asteroid. We found that the pyroxene chemistry for both asteroids is 
Fs$_{40.6\pm3.3}$Wo$_{8.9\pm1.1}$; these values fall within the range of HED meteorites \citep[e.g.][]{1998LPI....29.1220M} and are similar to the ones calculated for some M-type 
asteroids in the main 
belt \citep[e.g.][]{2011M&PS...46.1910H}. 

\cite{2017AJ....153...29S} found a correlation between the Band I depth (BD) and the pyroxene abundance in intimate mixtures of orthopyroxene and metal. This correlation is described by the following second-order polynomial 
fit:

\begin{equation}
opx/(opx+metal) = -0.000274\times BD^{2}+0.033\times BD+0.014
\end{equation}

where $opx/(opx + metal)$ is the orthopyroxene-metal abundance ratio. Using the measured Band I depth and this spectral calibration, we estimated an orthopyroxene abundance of 0.15$\pm$0.01 for both asteroids. This value is more 
than twice the mean value calculated for (16) Psyche by \cite{2017AJ....153...29S}.

\section{Source Region and Parent Body}

\begin{figure*}[!ht]
\begin{center}
\includegraphics[height=9cm]{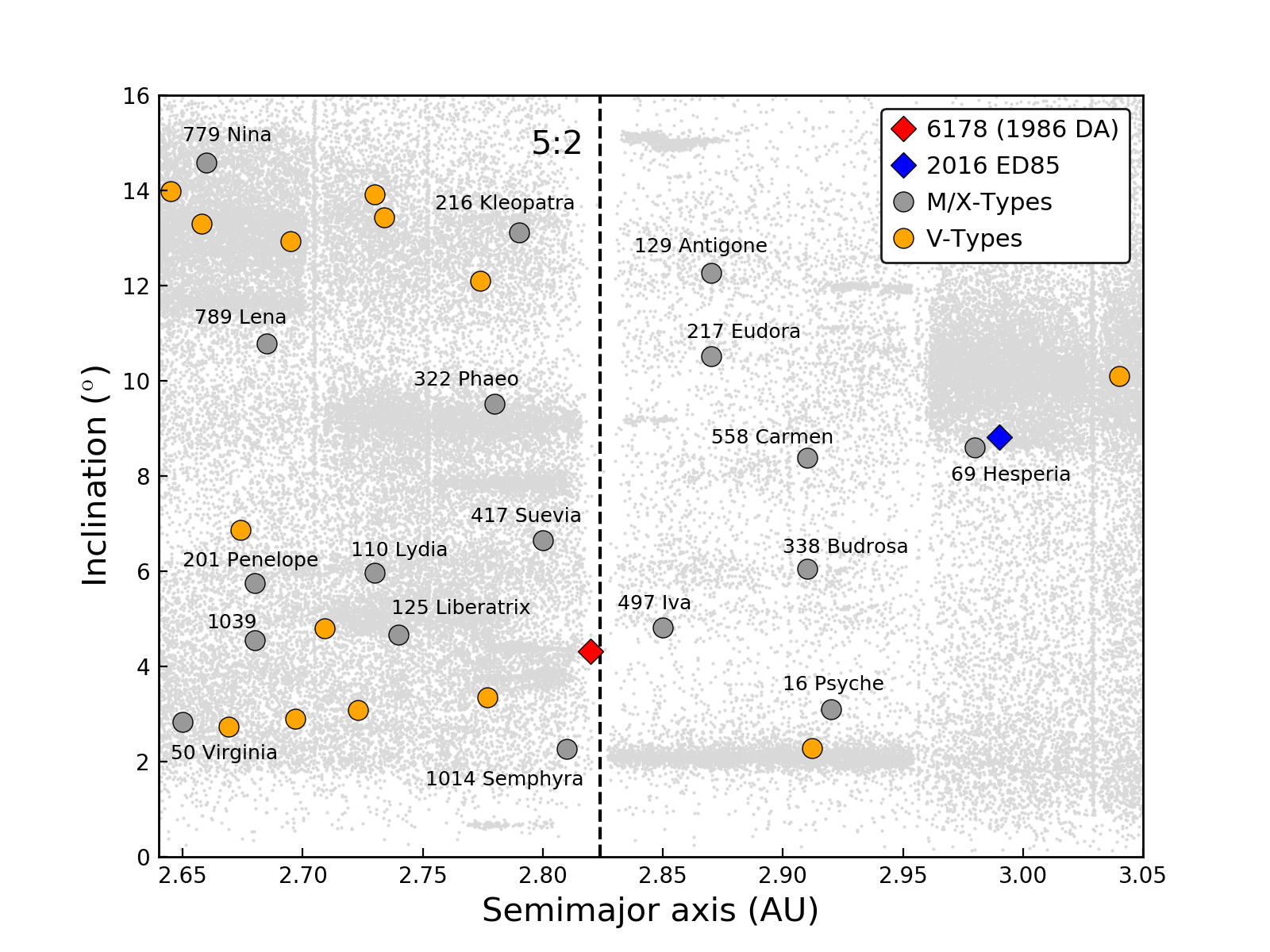}

\caption{\label{f:Figure3} {\small Inclination vs. semimajor axis for 6178 (1986 DA), 2016 ED85, and known M/X-types \citep[e.g.,][]{2004AJ....128.3070C, 2010Icar..208..221S, 2010Icar..210..674O, 2011M&PS...46.1910H, 2014Icar..238...37N}, and V-types \citep{2017Icar..295...61L, 2018AJ....156...11H, 2020MNRAS.491.5966M} in the middle and outer belt. The number and name of the M/X-type asteroids are indicated. The location of the 5:2 mean motion resonance is 
indicated with a vertical dashed line. Background asteroids are depicted in light gray and were obtained from the Asteroids Dynamic Site (AstDyS-2). For clarity, only background asteroids with $H$ $<$ 16 are shown.}}

\end{center}
\end{figure*}

The orbits of 1986 DA and 2016 ED85, with semimajor axes near the 5:2 mean motion resonance with Jupiter (5:2 MMR) or beyond, are suggestive that they originated in the outer asteroid belt. This is 
also the region where some of the largest known metal-rich asteroids reside. Due to their particular location, it is plausible that 1986 DA  and 2016 ED85 (if confirmed to be metal-rich) are fragments from such bodies. The location of several 
known M/X-type asteroids is shown in Figure \ref{f:Figure3}. It is worth mentioning that both 1986 DA and 2016 ED85 are now on planet-crossing orbits, and they reached those locations by entering into resonances that change 
their inclination values. For that reason, Figure \ref{f:Figure3} should be interpreted with caution; a similarity in inclination between a given M/X-type object and either of our NEAs may be a coincidence.

In order to make a quantitative assessment of likely source regions for these bodies, we input them into the near-Earth-object (NEO) model described by \cite{2017A&A...598A..52G, 2018Icar..312..181G}. In their model, over 
70,000 test asteroids with diameter $D$ = 0.1 and 1 km were dynamical tracked as they escaped the main asteroid belt and transneptunian populations. These test asteroids were followed until they hit a planet, the Sun, or were ejected 
out of the inner solar system via a close encounter with Jupiter. Most of the main belt bodies escaped out one of several regions, including the $\nu_{6}$ secular resonance, the 3:1, 5:2, and 2:1 MMR with Jupiter; and the Jupiter family 
comet region. From here, residence time probability distributions were created that calculated how much time these bodies spent in different (a, e, i) bins. By summing all of the probability distributions together with different weighting 
values, and then combining them with an observational bias model for the detection of NEOs, they were able to fit their NEO model population to observed NEOs. By varying the weighting values, they found that a best fit model yields not only an estimate of the debiased NEO orbital distribution but also the relative importance that a given NEO source region provides objects to each (a, e, i) bin. Hence, by inputing the (a, e, i) orbits of 1986 DA and 2016 
ED85 into this model, we can make predictions of their probable source and departure location.    

We found that the most likely region from which 1986 DA and 2016 ED85 originated is the 5:2 MMR with Jupiter near 2.8 au, with probabilities of 76\% and 49\%, respectively (Table 2). We argue that the probabilities make sense 
because there are numerous large M- and X-type asteroids residing near the borders of the 5:2 resonance (Figure \ref{f:Figure3}); several of them are plausible parent bodies for either NEA.

Now, it is possible that both 1986 DA and 2016 ED85 are objects from the background population of the asteroid belt. While both are favored to escape the 5:2 MMR, we also cannot rule out the possibility that they came from 
different NEO source regions, namely, those with lower but nonzero probabilities of reaching their listed (a, e, i) orbits.  It is also possible that the NEOs had different parent bodies. For this exercise, however, we will assume that the following is 
true: (i) the similarity in spectra between the objects is not a fluke, and they are connected to a single parent body or parent family; (ii) both came from the 5:2 MMR, the highest probability source in Table 2; and (iii) that the parent body or 
parent family can deliver $D$ $>$ 3 km bodies to the 5:2 MMR right now, a condition needed to explain both bodies given that their dynamical lifetimes are both of the order of a few Myr at best \citep{2002Icar..156..399B, 2018Icar..312..181G}.

\begin{table}[h]
\caption{\label{t:Table2} {\small Orbital parameters and possible regions from which 1986 DA and 2016 ED85 originated. The columns in this table are: object designation, semimajor axis (a), eccentricity (e), inclination (i), and probability 
(P) of having originated from the resonances $\nu_{6}$, 3:1, 5:2, 2:1, and the Jupiter family comet region (JFC).}}
\begin{tabular}{ccccccccc}
\tableline

Asteroid&	a (au)&e&	i ($^{\circ}$)&P ($\nu_{6}$)&P (3:1)&	P (5:2)&P (2:1)&P (JFC) \\  \hline
6178 (1986 DA)&2.82&0.58&4.31&0.02&0.08&	0.76&0.02&0.13 \\
2016 ED85&3.00&0.69&8.81&0.01&	0.04&0.49&0.11&0.36 \\

\tableline
\end{tabular}
\end{table}

\begin{figure*}[!ht]
\begin{center}
\includegraphics[height=9cm]{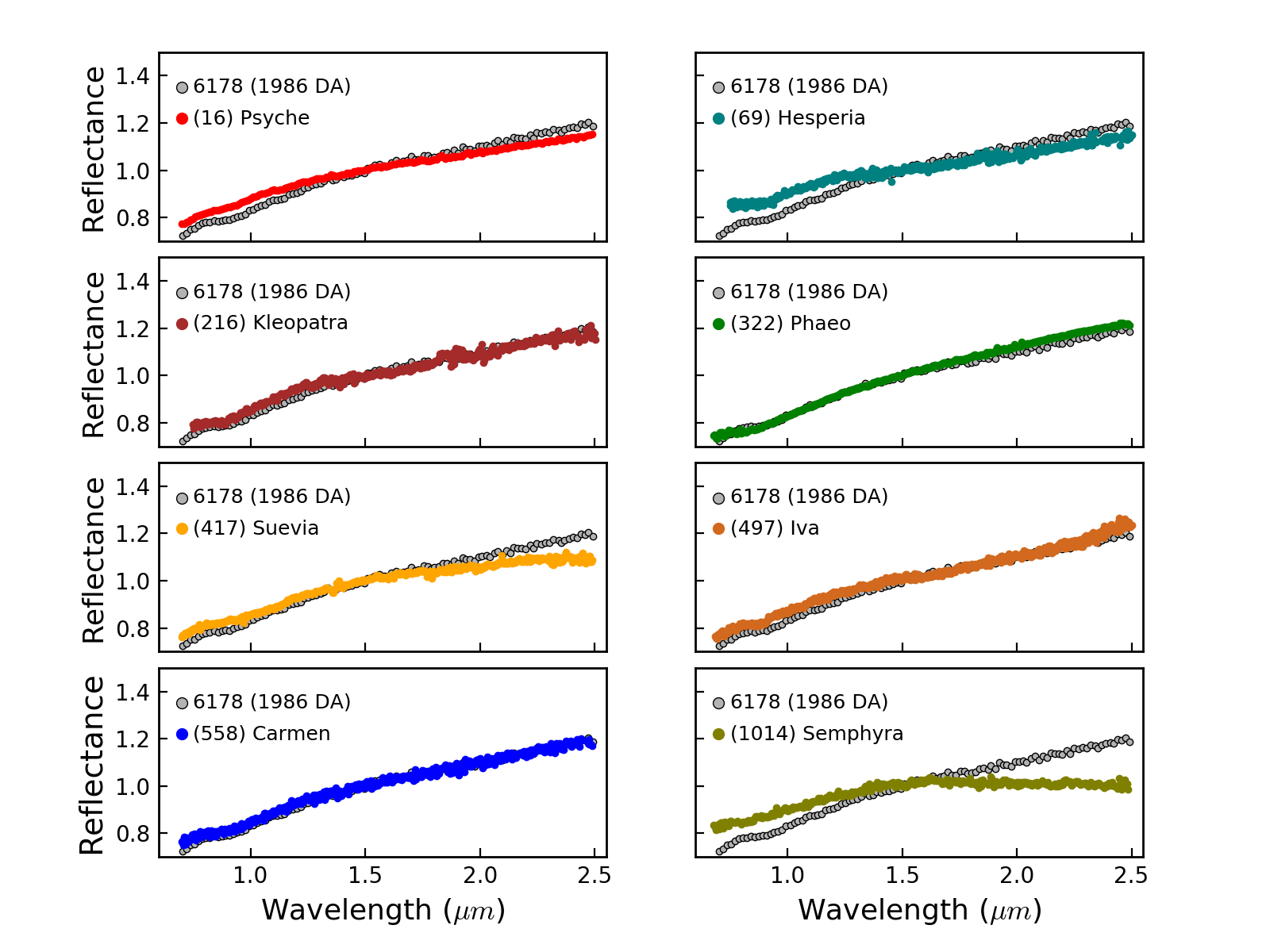}

\caption{\label{f:Figure4} {\small NIR spectra of 6178 (1986 DA) and M/X-type asteroids (16) Psyche \citep{2017AJ....153...29S}, (69) Hesperia \citep{2016PDSS..248.....H}, (216) Kleopatra \citep{2016PDSS..248.....H}, (322) Phaeo 
\citep{2004AJ....128.3070C}, (417) Suevia \citep{2016PDSS..248.....H}, (497) Iva (this work), (558) Carmen \citep{2016PDSS..248.....H}, and (1014) Semphyra \citep{2004AJ....128.3070C}. The spectrum of (497) Iva was obtained with the 
IRTF on January 02, 2020 UTC as part of this study. All spectra are normalized to unity at 1.5 $\mu$m.}}

\end{center}
\end{figure*}

\begin{figure*}[!ht]
\begin{center}
\includegraphics[height=9cm]{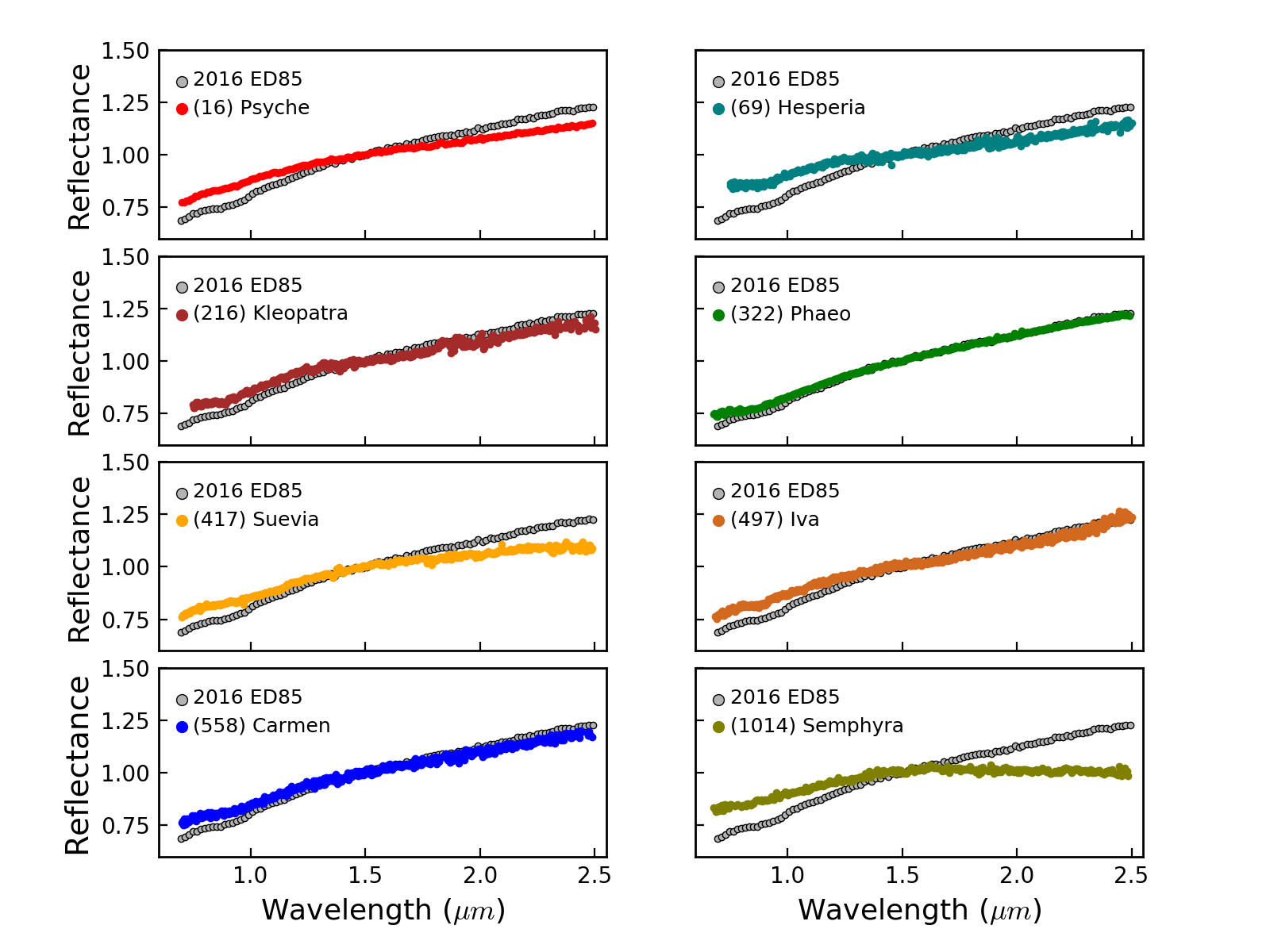}

\caption{\label{f:Figure5} {\small NIR spectra of 2016 ED85 and M/X-type asteroids (16) Psyche \citep{2017AJ....153...29S}, (69) Hesperia \citep{2016PDSS..248.....H}, (216) Kleopatra \citep{2016PDSS..248.....H}, (322) Phaeo 
\citep{2004AJ....128.3070C}, (417) Suevia \citep{2016PDSS..248.....H}, (497) Iva (this work), (558) Carmen \citep{2016PDSS..248.....H}, and (1014) Semphyra \citep{2004AJ....128.3070C}. The spectrum of (497) Iva was obtained with the 
IRTF on January 02, 2020 UTC as part of this study. All spectra are normalized to unity at 1.5 $\mu$m.}}

\end{center}
\end{figure*}

In order to separate the primary parent body candidates from the secondary ones, we compared the NIR spectra of 1986 DA and 2016 ED85 with the spectra of some of the M/X-type asteroids in the vicinity 
(Figures \ref{f:Figure4} and \ref{f:Figure5}). We found that the spectral characteristics of these two NEAs are consistent with some of the M/X-type asteroids in the main belt, which also exhibit red spectral slopes and a weak pyroxene 
absorption band at $\sim$ 0.9 $\mu$m. Overall, the candidate parent bodies with the most similar spectra were (216) Kleopatra, (322) Phaeo, (497) Iva, and (558) Carmen.

One interesting aspect about the metal-rich asteroids in the middle/outer belt is that few of them are associated with an asteroid family. If 1986 DA and 2016 ED85 are fragments that resulted from a collision between a metal-rich 
asteroid and another object, one would expect more fragments to be created after the collision, leaving behind an asteroid family composed of this type of object. In particular, to explain 1986 DA, which is roughly 3 km in diameter, a 
candidate family needs to be located adjacent to the 5:2 resonance and be truncated at sizes that are at least 3 km.  

\begin{figure*}[!h]
\begin{center}
\includegraphics[height=9cm]{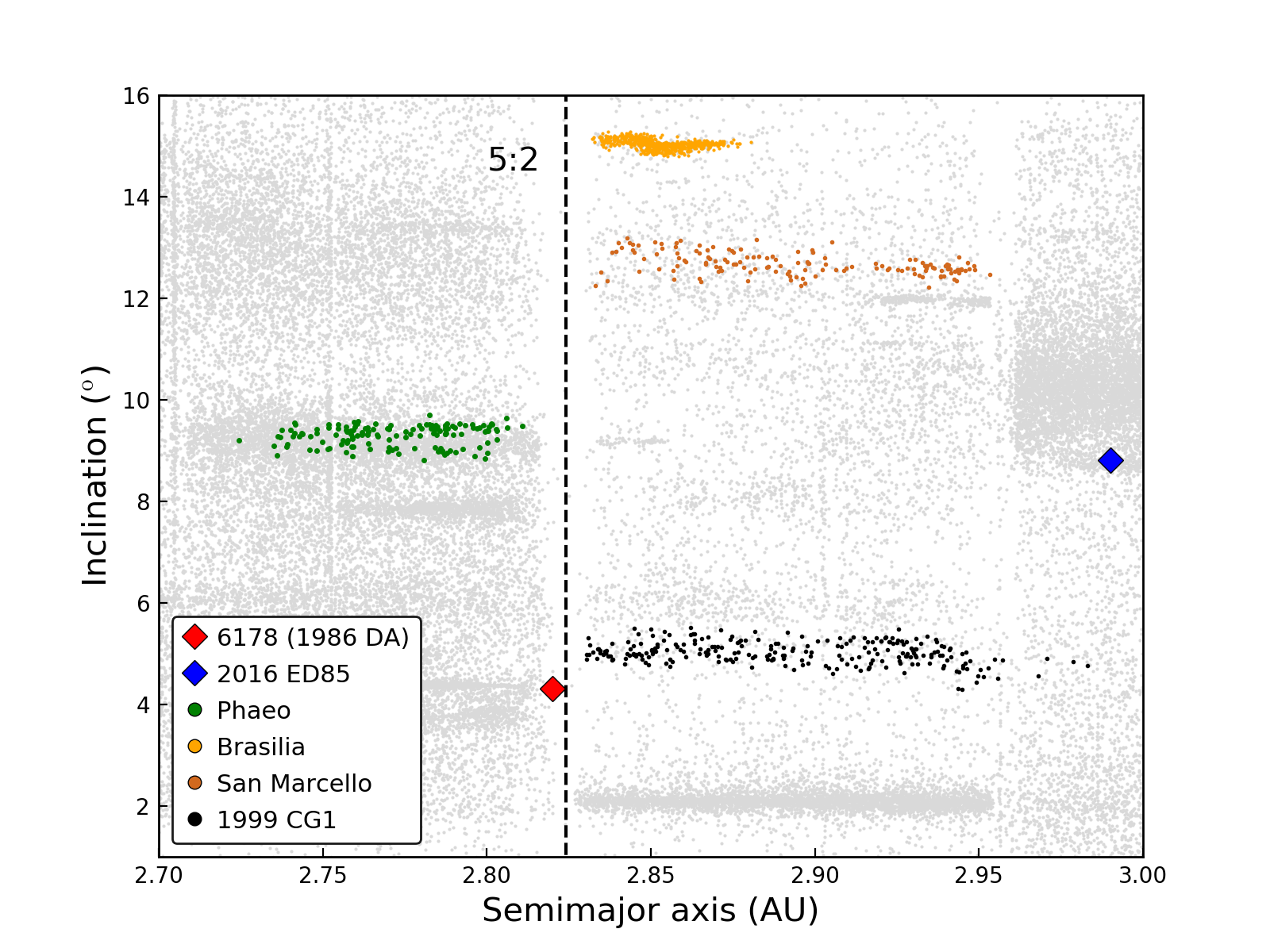}

\caption{\label{f:Figure6} {\small Inclination vs. semimajor axis for 6178 (1986 DA), 2016 ED85, and asteroid families Phaeo, Brasilia, San Marcello, and 1999 CG1 from \cite{2015PDSS..234.....N}. The location of the 5:2 mean motion 
resonance is indicated with a vertical dashed line. Background asteroids are depicted in light gray and were obtained from the Asteroids Dynamic Site (AstDyS-2). For clarity, only background asteroids with $H$ $<$ 16 are shown.}}

\end{center}
\end{figure*}

Many families have been identified in the middle and outer belt that are also close to the 5:2 resonance \citep{2015aste.book..297N}. Most of them can be ruled out on the basis of their albedo and spectroscopic data. However, there 
are four families that are worth mentioning: Phaeo, Brasilia, San Marcello, and 1999 CG1. Their proper semimajor axes vs. proper inclinations are shown in Figure \ref{f:Figure6}.

\textbf{Phaeo family}. To date, the only family in this region whose parent body has been found to have spectral characteristics similar to a metal-rich asteroid is the Phaeo family \citep{2004AJ....128.3070C}. It also fits the 
dynamical and size characteristics necessary to be a source of both NEAs. Using test bodies evolving from orbits similar to those in the Phaeo family, we find that many can reach the present-day orbit of 1986 DA and 2016 ED85.

A possible problem with this family, however, is that the taxonomic classification of (322) Phaeo remains ambiguous, since this object was originally classified as a D-type by \cite{2009Icar..202..160D}. The albedo of (322) Phaeo 
\citep[0.0837$\pm$0.0178;][]{2012Icar..221..365P}, on the other hand, is fairly consistent within errors with the albedo of 1986 DA (0.096$\pm$0.029) derived by \cite{2014ApJ...785L...4H}, although we note that the mean albedo of the family, $\sim$ 0.06 \citep{2015aste.book..323M, 2015aste.book..297N}, appears to be lower than the typical values for metal-rich asteroids. More spectroscopic observations of (322) Phaeo and members of its family are required to determine whether this is indeed a metal-rich asteroid family.

\textbf{Brasilia family}. The Brasilia family is composed of X-type asteroids with a mean albedo of 0.18 \citep{2015aste.book..323M, 2015aste.book..297N}. \cite{2015aste.book..297N} noted that (293) Brasilia could be an interloper, in which case (1521) 
Seinajoki would be the largest asteroid of this family.  Although NIR spectroscopic data are not available, visible spectra of some members of this family 
show a red slope with the possible presence of a 0.9 $\mu$m band \citep{2002Icar..158..106B}, consistent with metal-rich asteroids. The family’s albedo, however, is higher than that of 1986 DA.

\textbf{San Marcello family}. The San Marcello family shares similar characteristics with the Brasilia family, i.e., it is composed of X-type asteroids with a mean albedo of 0.19 \citep{2015aste.book..323M, 2015aste.book..297N}. The proximity of this family to 
the resonance makes it a possible candidate; however, NIR spectroscopic data are required to confirm its affinity with metal-rich asteroids.  

\textbf{1999 CG1 family}. According to \cite{2015aste.book..297N}, this family is composed of S-type asteroids. The reason why it has been included in our list is because the mean albedo of the family ($\sim$0.10) is too low for S-type 
asteroids but within the range of metal-rich asteroids and 1986 DA.  This could make it a plausible fit.  At this time, there are no NIR spectroscopic data available that could be used to confirm its composition. 

\begin{figure*}[!ht]
\begin{center}
\includegraphics[height=10cm]{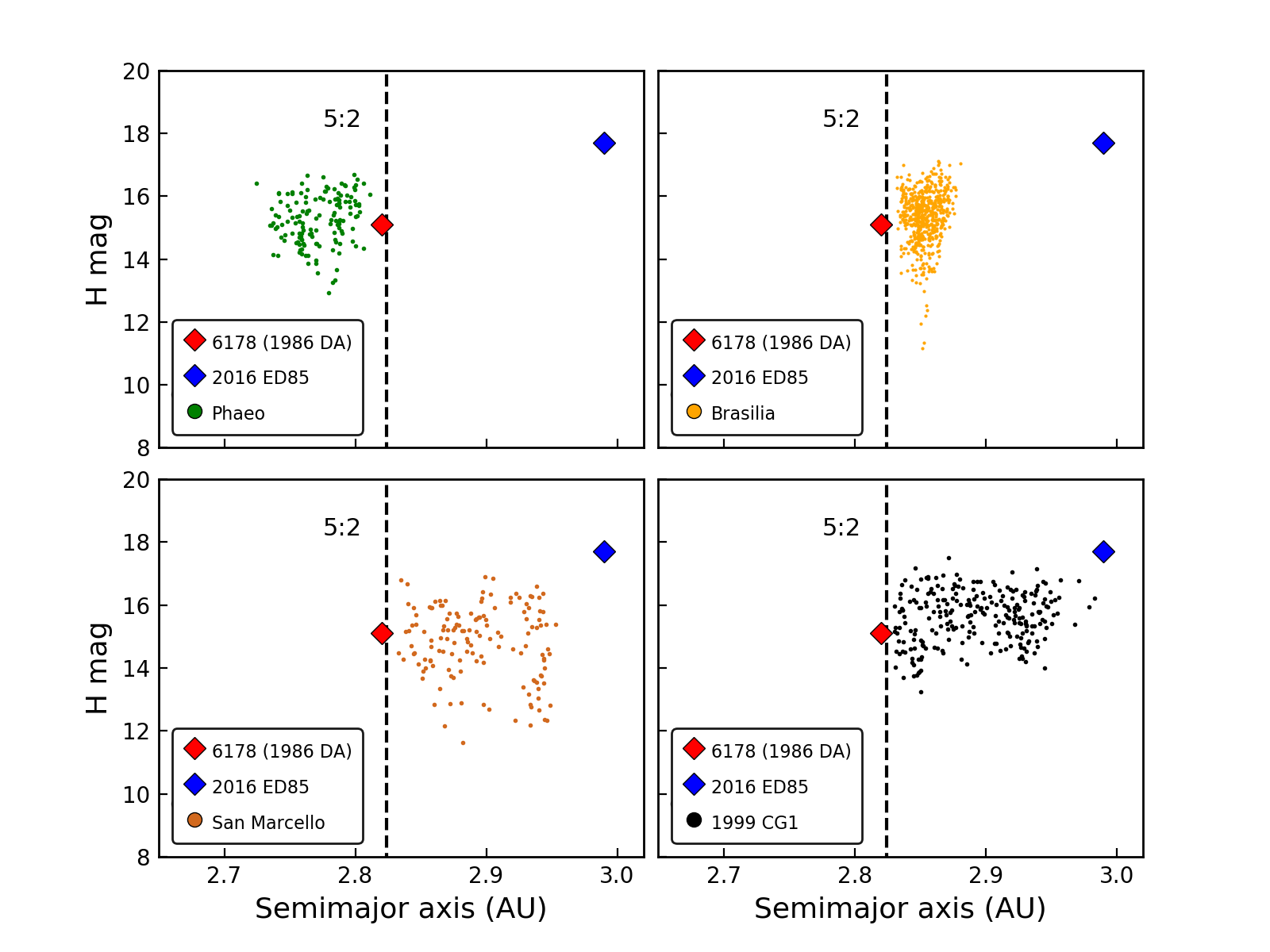}

\caption{\label{f:Figure7} {\small Absolute magnitude vs. semimajor axis for 6178 (1986 DA), 2016 ED85, and asteroid families Phaeo, Brasilia, San Marcello, and 1999 CG1 from \cite{2015PDSS..234.....N}. The location of the 5:2 mean motion resonance is indicated with a vertical dashed line.}}

\end{center}
\end{figure*}

In addition to the fact that these families border the 5:2 resonance, it is also important to look at the expected contribution rates of resonant objects in the size range of 2016 ED85 and 1986 DA. In Figure \ref{f:Figure7} we plot the 
absolute magnitude vs. semimajor axis for the families described above. All of them show the characteristic V shape resulting from the size-dependent semimajor axis drift due to the Yarkovsky effect. The wider V shape of the San 
Marcello and 1999 CG1 families indicates their older age compared to the other two families. All the families seem to have produced resonant objects in the size range of 1986 DA and can produce asteroids of the size of 2016 ED85; 
they have observed objects that are truncated by the 5:2 resonance, and that implies that bodies too small to be detected have also reached the same resonance. 

In summary, in this exercise we have considered the scenario in which the two NEAs come from the same parent body and are fragments from an asteroid family. Although the limited data prevent 
us from doing a more rigorous analysis, we have identified the Phaeo family as a candidate to produce 1986 DA and 2016 ED85 under this scenario. The family has the following:

\begin{itemize}

\item	A large remnant (322 Phaeo) whose spectroscopic signature is similar to both NEAs.
\item	A parent body with an albedo that is consistent with the albedo of 1986 DA.
\item	An orbital distribution that shows that the family is adjacent to the 5:2 resonance.
\item	Orbital evidence from the family that it has been delivering 1986 DA-sized and smaller bodies to the 5:2 resonance.
\item	Test bodies that can reach the present-day orbits of both NEAs according to the dynamical test body runs provided by \cite{2017A&A...598A..52G, 2018Icar..312..181G}. 

 \end{itemize}

The main issue about this family is that it could be composed of primitive bodies, given its low albedo and the fact that (322) Phaeo has been classified as a D-type asteroid. D-types have 
spectra with very steep slopes but are featureless in the NIR \citep{2009Icar..202..160D}. The detection of the 0.9 $\mu$m band in the NIR spectrum of (322) Phaeo by \cite{2004AJ....128.3070C} 
would rule out the D-type taxonomy for this object; however, only one NIR spectrum of (322) Phaeo has been analyzed. More spectroscopic data of this asteroid and members of its family would 
help to confirm their taxonomic type and constrain their composition. The other families, Brasilia, San Marcello, and 1999 CG1, are considered secondary candidates. Where we have 
information, they appear to have different spectroscopic signatures and albedos from either of our NEAs. We hesitate to rule them out, though, for the reasons listed above.

\section{Meteorite Analogs}

The next step in this study is to identify possible meteorite analogs for 1986 DA and 2016 ED85. Because 1986 DA is a mixture of rocks and metal, and there is the possibility for 2016 ED85 to be it as well, we look at 
stony-iron meteorites and metal-rich carbonaceous chondrites as possible analogs, since they share similar characteristics. Enstatite chondrites, which have been considered as meteorite analogs for metal-rich asteroids, are not considered 
here because the pyroxene in these meteorites is iron-free and does not exhibit the 0.9 $\mu$m feature present in the spectra of the NEAs. Stony-iron meteorites that have been suggested as potential analogs for metal-rich asteroids include pallasites and mesosiderites, whereas metal-rich carbonaceous chondrites include high metal carbonaceous chondrites (CH) and bencubbinite (CB) chondrites.

Pallasites consist mostly of metal and olivine in roughly equal amounts, with troilite as a minor phase \citep{1998LPI....29.1220M}. They are thought to be fragments from the core-mantle boundary of differentiated asteroids. We have 
ruled out pallasites as meteorite analogs because the presence of olivine would produce an absorption feature centered at $\sim$ 1.05 $\mu$m, which would not match the Band I center measured for 1986 DA and 2016 ED85 at 
0.93 $\mu$m. 

Mesosiderites are breccias composed of similar proportions of silicates and FeNi metal. The silicate component is very similar in composition to HED meteorites, containing pyroxene, olivine, and Ca-rich feldspar 
\citep[e.g.,][]{1979LPSC...10.1109H, 1998LPI....29.1220M, 2001M&PS...36..869S}. Two models have been proposed to explain the origin of mesosiderites. In the first model, the metal and the silicates would have originated in two different bodies following the collision between a metallic core and the basaltic surface 
of a differentiated asteroid \citep{1985Natur.318..168W}. In the second model, both the metal and the silicates originate in a single differentiated asteroid with a molten core and are mixed together after the asteroid is disrupted by 
another object \citep{2001M&PS...36..869S}. \cite{1993Icar..101..201R} found it unlikely that mesosiderites and HEDs formed in the same parent body, although a more recent study by \cite{2019NatGe..12..510H} points to asteroid (4) Vesta as the parent body of both meteorites. Mesosiderites have also been linked to the Maria asteroid family located adjacent to the 3:1 MMR \citep{2011Icar..213..524F}.

The CH chondrites are polymict breccias characterized by the presence of small cryptocrystalline chondrules and a high abundance of Fe,Ni-metal ($\sim$20 vol\%) 
\citep{2006mess.book...19W}. These meteorites are believed to have formed in the solar nebula \citep[e.g.,][]{1988E&PSL..91...19W, 2004GeCoA..68.3409C}. Their 
bulk density is higher than most carbonaceous chondrites but lower than that of objects like 1986 DA, suggesting that CH chondrites are not good meteorite analogs.

Bencubbinites are also breccias containing high metal abundances of $\sim$ 60-80 vol\% and chemically reduced silicates including Fe-poor olivine, low-Ca pyroxene, and high-Ca pyroxene. Calcium–aluminum-rich inclusion are also 
found in some of these meteorites \citep{1998M&PSA..33Q.166W, 2001M&PS...36..401W}. They are thought to have formed either directly from the solar nebula \citep[e.g.,][]{1979GeCoA..43..689N, 1990Metic..25..269W, 2001M&PS...36..401W} or in a metal-enriched gas resulting from a protoplanetary impact \citep[e.g.,][]{2002GeCoA..66..647C, 2005Natur.436..989K}. 

In this work we focused on the analysis of mesosiderites and bencubbinites. Despite the fact that these meteorites have been considered as possible analogs for metal-rich asteroids, powder spectra are scarce, making it difficult to establish a linkage between them and the 
asteroids. For this study we used two samples, the mesosiderite NWA 6370 and the bencubbinite Gujba; both samples have similar proportions of silicates and metal. The 
samples were crushed with a pestle and mortar and dry sieved to three different grain sizes: $<$45, $<$150, and 150-500 $\mu$m. This broad range of grain sizes was chosen to investigate whether increasing the grain size could 
produce a better fit between the meteorite and asteroid spectra. Visible and NIR spectra (0.35-2.5 $\mu$m) were obtained relative to a Labsphere Spectralon disk using an ASD Labspec4 Pro spectrometer at an incident angle 
$i$=0$^{\circ}$ and emission angle $e$=30$^{\circ}$. For each measurement, 1000 scans were obtained and averaged to create the final spectrum. Spectral band parameters were measured for all samples in the same way as was done for 
the asteroid spectra; the measured values are presented in Table 3.

\begin{table}[h]
\caption{\label{t:Table3} {\small Spectral band parameters for the mesosiderite NWA 6370 and the bencubbinite Gujba. The columns in this table are: sample, grain size, albedo (reflectance value measured at 0.55 $\mu$m), Band I 
center (BIC), Band II center (BIIC), Band I depth (BID), and Band II depth (BIID).}}
\begin{tabular}{ccccccc}
\tableline

Sample&Grain Size ($\mu$m)&Albedo&BIC ($\mu$m)&BIIC ($\mu$m)&BID (\%)&BIID (\%) \\  \hline

NWA 6370&$<$45&0.11&0.929$\pm$0.001&1.929$\pm$0.004&18.4$\pm$0.1&10.6$\pm$0.2 \\
NWA 6370&$<$150&0.14&0.932$\pm$0.001&1.953$\pm$0.005&30.7$\pm$0.1&22.7$\pm$0.2 \\
NWA 6370&150-500&0.09&0.930$\pm$0.001&1.953$\pm$0.007&12.1$\pm$0.1&7.8$\pm$0.1 \\
Gujba&$<$45& 0.16&0.918$\pm$0.002&-&3.8$\pm$0.1&- \\
Gujba&$<$150&0.14&0.920$\pm$0.003&-&5.3$\pm$0.1&- \\
Gujba&150-500& 0.08&0.919$\pm$0.003&-&7.7$\pm$0.2&- \\

\tableline
\end{tabular}
\end{table}

\begin{figure*}[!h]
\begin{center}
\includegraphics[height=10cm]{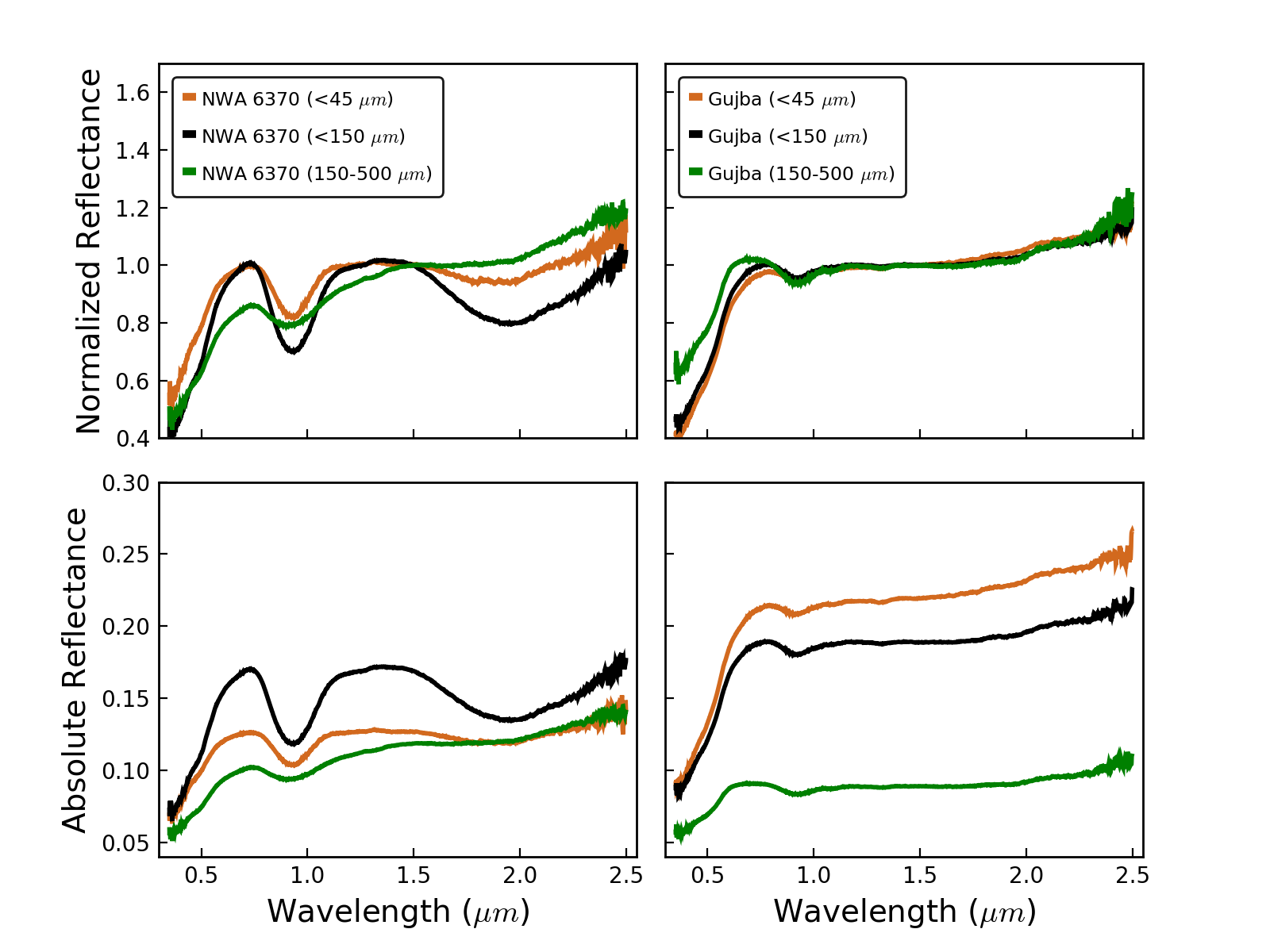}

\caption{\label{f:Figure8} {\small Top row: normalized reflectance for the mesosiderite NWA 6370 (left) and the bencubbinite Gujba (right) for three different grain sizes. All the spectra are normalized to unity at 1.5 $\mu$m. Bottom 
row: absolute reflectance for three different grain sizes for the mesosiderite NWA 6370 (left) and the bencubbinite Gujba (right).}}

\end{center}
\end{figure*}

\begin{figure*}[!h]
\begin{center}
\includegraphics[height=9cm]{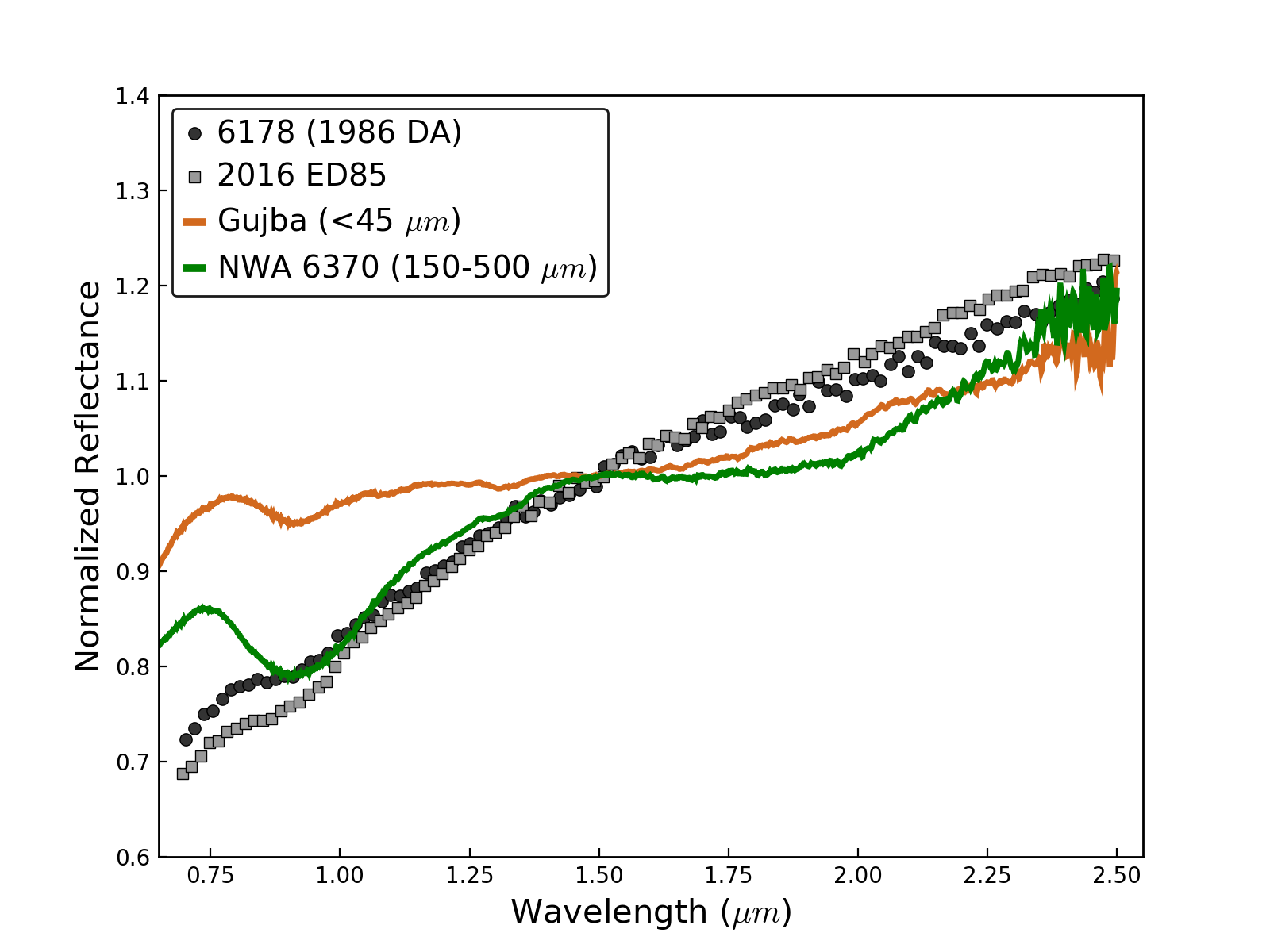}

\caption{\label{f:Figure9} {\small Spectra of 6178 (1986 DA) and 2016 ED85 compared with the spectra of the bencubbinite Gujba ($<$45 $\mu$m) and the mesosiderite NWA 6370 (150-500 $\mu$m). The spectra have been 
normalized to unity at 1.5 $\mu$m.}}

\end{center}
\end{figure*}

Figure \ref{f:Figure8} shows the normalized spectra (top row) for the two samples. All the mesosiderite spectra exhibit two absorption bands centered at $\sim$0.93 and 1.94 $\mu$m.  Variations in the intensity of the absorption 
bands with grain size are evident for the mesosiderite NWA 6370. The spectrum corresponding to the largest grain size exhibits the weakest absorption bands and the steepest spectral slope, producing the closest match in terms of 
spectral slope with the asteroid spectra (Figure \ref{f:Figure9}). However, these large grains will also cause more light to be absorbed, resulting in a significant decrease in reflectance for this spectrum. This can be seen in Figure 
\ref{f:Figure8} (bottom row), where the absolute reflectances for the three grain sizes are compared. The bencubbinite spectra only show one absorption band centered at $\sim$ 0.92 $\mu$m. For Gujba, we also observed variations in the 
band depth, although not as pronounced as NWA 6370. All the spectra have a relatively flat slope; the spectrum corresponding to the grain size of $<$ 45 $\mu$m has a slightly steeper slope than the others, but much less than the 
asteroid spectra (Figure \ref{f:Figure9}). Like NWA 6370, the largest grain size shows the lowest reflectance.

The use of different grain sizes did not help to improve the spectral match between the 
bencubbinite and the asteroid spectra. In the case of the mesosiderite, the largest grain size produced a better fit of the spectral slope; however, the absorption bands are still too deep to produce a good match with the asteroid spectra. 
These results indicate that grain size alone is not responsible for the differences observed between the meteorite and the asteroid spectra. 

The compositional analysis of 1986 DA and 2016 ED85 showed that both objects have a pyroxene chemistry comparable to that of HEDs, but inconsistent with the magnesian pyroxene found in bencubbinites. This difference in the 
pyroxene chemistry and the significant spectral differences found in this work suggest that bencubbinites are not good meteorite analogs for 1986 DA and 2016 ED85. The situation with the mesosiderites is less clear; the pyroxene chemistry
for some of these meteorites is consistent with the values found for the two NEAs, but their 
spectra have deeper absorption bands and less pronounced spectral slopes. One possible explanation for these spectral differences could be the effect of space weathering on the NEAs, which is known to produce reddening of the spectral slope and suppression of the absorption bands
\citep[e.g.,][]{2001JGR...10610039H, 2006Icar..184..327B, 2010Icar..209..564G}. However, irradiation experiments performed on the mesosiderite Vaca Muerta by \cite{2009Icar..202..477V} showed that space weathering is 
not enough to reproduce the red spectral slopes of metal-rich asteroids.

\begin{figure*}[!ht]
\begin{center}
\includegraphics[height=8cm]{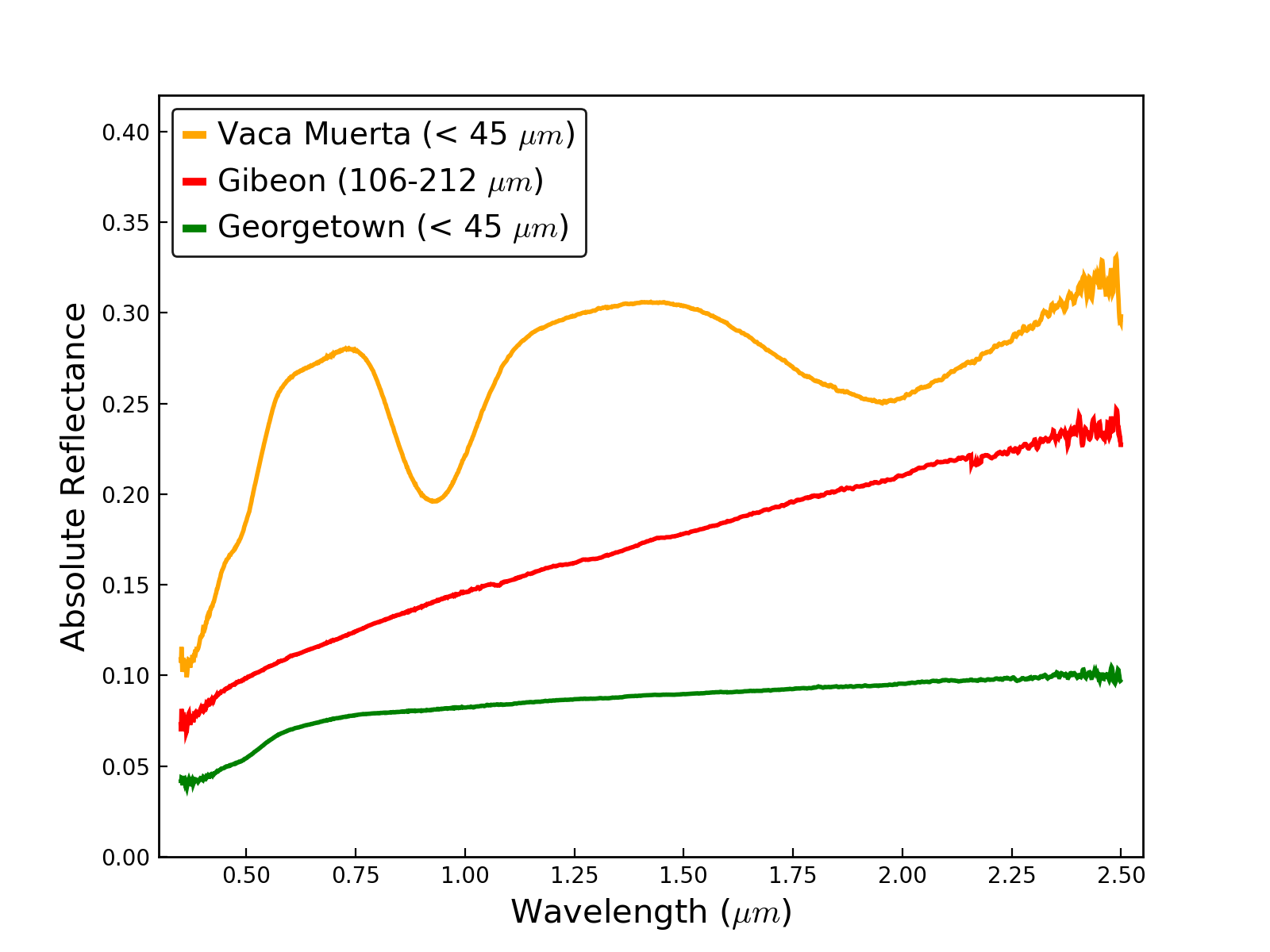}

\caption{\label{f:Figure10} {\small Spectra of the endmembers used with the mixing model. The samples include the mesosiderite Vaca Muerta, the iron IVA meteorite Gibeon, and the iron IAB meteorite Georgetown. The grain 
size for each sample is indicated.}}

\end{center}
\end{figure*}

If grain size and space weathering cannot explain the spectral differences between the mesosiderites and the metal-rich asteroids, perhaps metal abundance could, since adding metal would increase the spectral slope and suppress 
the absorption bands. To explore this possibility, we modeled the spectra of 1986 DA and 2016 ED85 with a mixing model \citep{2001JGR...10610039H} that allowed us to combine the silicate component of a mesosiderite with meteoritic metal. For this study we used the spectrum of the mesosiderite Vaca Muerta ($<$45 $\mu$m). This sample was used instead of NWA 6370 because it is composed almost entirely of pyroxene with little metal, which is 
preferred in this case in order to have a more accurate estimate of the amount of metal present in the mixture. The sample was prepared in the same way as NWA 6370, and the spectra were obtained under the same configuration. For the metal 
component we used the spectra of two iron meteorites, Gibeon (IVA) and Georgetown (IAB). The Gibeon powder was obtained as cutting shavings, which were sieved with acetone to grain sizes of 106-212 $\mu$m and dried in a 
heated vacuum oven. The Georgetown powder was obtained by crushing a small fragment with a pestle and mortar, and the powder was then dry sieved to a grain size of $<$45 $\mu$m. These are the 
same samples used in \cite{2021PSJ.....2...95C}. The spectra were obtained following the same procedure as used with NWA 6370 and Gujba. All the endmember spectra are shown in Figure \ref{f:Figure10}.

A combination of three endmembers was used for the mixing model, the two iron meteorites plus the mesosiderite. Using two iron meteorites whose spectra have different spectral slopes has the advantage of allowing a better fit of the 
spectral slope. Asteroid spectra were extrapolated and normalized to unity at 0.55 $\mu$m. The normalized spectra were then multiplied by their geometric albedos in order to convert from relative reflectance to absolute reflectance. For 1986 DA we used the revised value $P_{V}$=0.096$\pm$0.029 obtained by \cite{2014ApJ...785L...4H}. Since the albedo of 2016 
ED85 is not known, we assumed the same value calculated for 1986 DA. Although 2016 ED85 can only be considered a candidate metal-rich body at this point, 
we wanted to see whether it was possible to model its NIR spectrum in the same way as 1986 DA.

From \cite{2001JGR...10610039H}, the reflectance relative to a standard can be written as

\begin{equation}
\Gamma(\gamma)=\frac{r(sample)}{r(standard)}=\frac{1-\gamma^{2}}{(1+2\gamma \mu_{0})(1+2\gamma \mu)}
\end{equation}

where $\mu_{0}$=$cos(i)$ and $\mu$=$cos(e)$. From Equation (5), $\gamma$ can be determined and the single scattering albedo for each endmember calculated as
   
\begin{equation}
w=1-\gamma^{2}
\end{equation}

The $w$ values are then linearly combined and converted to reflectance with Equation (5). This was done with a Python routine that incorporates the {\it{curve\_fit}} function included in the SciPy library for Python 
\citep[e.g.,][]{2020NatMe..17..261V}. This function uses the Levenberg-Marquardt algorithm, which performs a nonlinear least-squares fit of the function to the data. The routine then returns the optimal quantities for the endmembers so 
that the sum of the squares of the differences between the function and the data ($\chi^{2}$) is minimized. The metal abundances of the two iron meteorites were left as free parameters, whereas the pyroxene contribution from Vaca 
Muerta was given a fixed value of 0.15 (see section 4). For each asteroid the albedo was allowed to vary within its error. Mineral abundances obtained from the mixing models are presented in Table 4.

\begin{table}[h!]
\begin{center}
\caption{\label{t:Table4} {\small Mineral abundances obtained from the mixing model. The albedo corresponds to the reflectance value measured at 0.55 $\mu$m for each fit.}}
\begin{tabular}{cccccc}
\tableline

Asteroid&Gibeon&Georgetown&Vaca Muerta&Albedo&$\chi^{2}$  \\  \hline
(16) Psyche&0.37&0.57&0.06&	0.088&0.001 \\
6178 (1986 DA)&0.57&0.28&0.15&0.105&0.003 \\
2016 ED85&0.70&0.15&0.15&0.114&0.004 \\

\tableline
\end{tabular}
\end{center}
\end{table}

\begin{figure*}[!ht]
\begin{center}
\includegraphics[height=8.5cm]{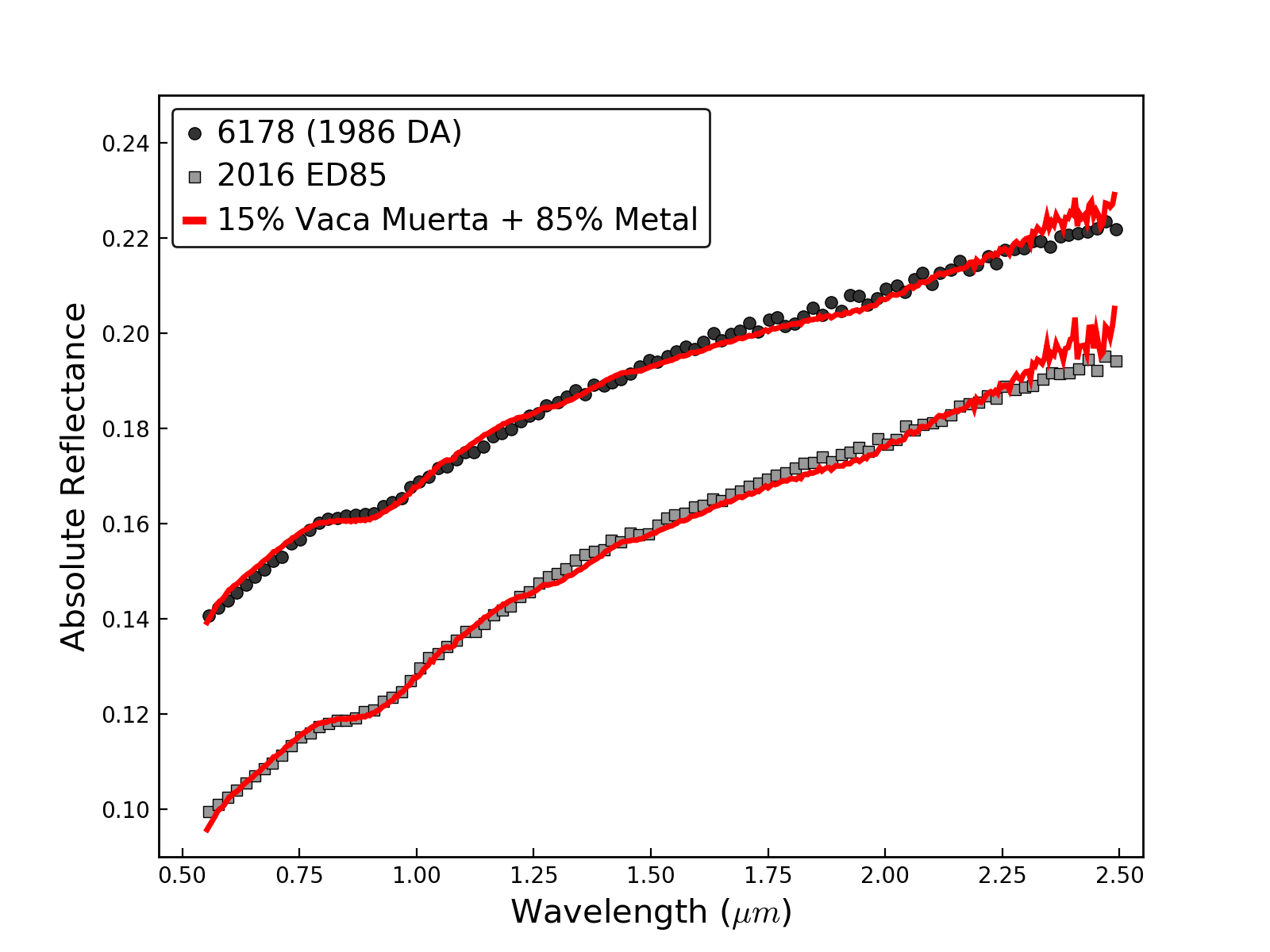}

\caption{\label{f:Figure11} {\small Spectra of 6178 (1986 DA) and 2016 ED85, and the best fit obtained using the mixing model. The silicate component comes from the mesosiderite Vaca Muerta, the metal component from the iron 
meteorites Gibeon and Georgetown. The spectra have been offset for clarity.}}

\end{center}
\end{figure*}

\begin{figure*}[!ht]
\begin{center}
\includegraphics[height=8.5cm]{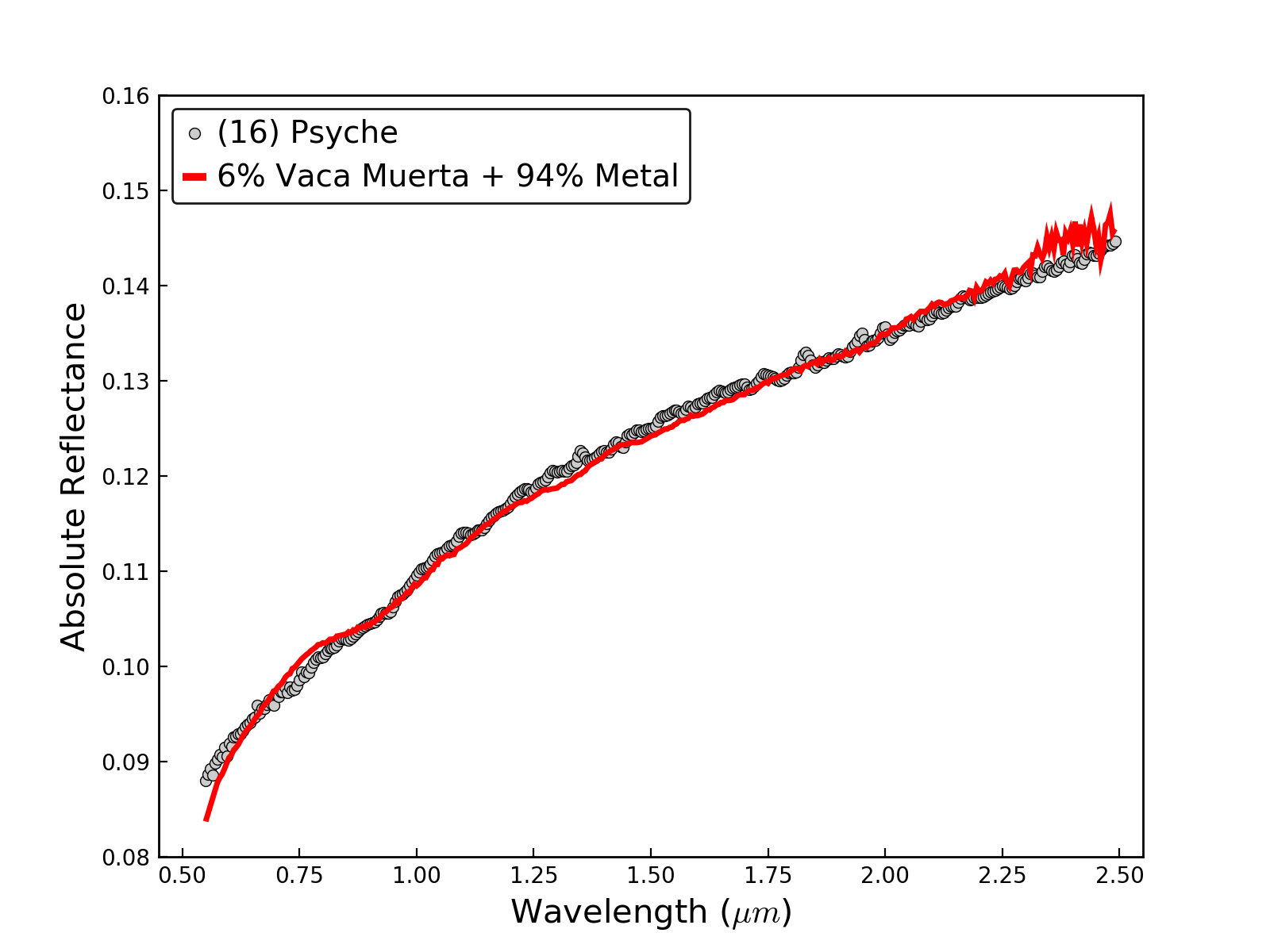}

\caption{\label{f:Figure12} {\small Spectrum of asteroid (16) Psyche from \cite{2017AJ....153...29S}, and the best fit obtained using the mixing model. The silicate component comes from the mesosiderite Vaca Muerta, the metal 
component from the iron meteorites Gibeon and Georgetown. The visible part of the spectrum was obtained from \cite{2002Icar..158..106B}.}}

\end{center}
\end{figure*}

Figure \ref{f:Figure11} shows the best fits obtained for 1986 DA and 2016 ED85. In general, we found that increasing the metal content significantly improves the match between the meteorite and the asteroid spectra. We performed the 
same analysis on the spectra of other metal-rich asteroids and obtained similar results. As an example, Figure \ref{f:Figure12} shows the best fit obtained for (16) Psyche; in this case, the pyroxene contribution from Vaca 
Muerta was fixed to a value of 0.06 based on the results obtained by \cite{2017AJ....153...29S}. For Psyche, the metal component from Georgetown contributed much more to the fit than for the NEAs, which required a higher quantity 
from Gibeon to match the spectral slope (Table 4). 

The results from the mixing model demonstrate that a higher metal content on the surface of the NEAs and other metal-rich asteroids could explain the spectral differences between them and the mesosiderites. As stated earlier, there is no 
consensus about the origin of these meteorites; if Vesta is their parent body, then we would have to rule out a link between the mesosiderites and the metal-rich bodies in the middle and outer belt. Despite this possibility, our results suggest that the parent body of asteroids 
like 1986 DA and possibly 2016 ED85 shares similar characteristics with the parent body of the mesosiderites, and that similar events that led to the formation of these meteorites could have taken place in other parts of the solar 
system. Evidence for this would be given by the presence of basaltic (V-type) asteroids not related to Vesta in the same region where some of the largest M/X-types reside between $\sim$2.65 and 3.0 au (Figure \ref{f:Figure3}). 

\section{1986 DA  as a target for asteroid mining}

The idea of asteroid mining is not new; \cite{1977TechnolRev....79...7G} discussed the benefits of mining extraterrestrial resources, not only for the potential economic value but also as a way of reducing the environmental damage on 
Earth. They considered the scenario of mining a small asteroid containing one cubic kilometer of Ni-Fe metals and estimated that for a delivery rate of 50,000 metric tons of nickel per day, the annual return at that time would have been \$100 
billion. \cite{1994JGR....9921129K, 1996USGS....2821K} argued that because of the abundance and low prices of Fe and Ni, their exploitation for use on Earth would not be required in the short term, although he pointed out that they 
could be used in space construction. Instead, he considered exploiting precious metals such as Au and the platinum group metals (PGM), which include Ru, Rh, Pd, Os, Ir, and Pt.  \cite{1996USGS....2821K} 
evaluated three mining scenarios, one of them involving the exploitation of a 2.6 km metallic NEA in the 90$^{th}$ percentile of Pt richness. He found that the annual value of precious metals from this object (in 1995 U.S dollars), if 
marketed over 50 yr, would be $\sim$ \$48 billion.  

Due to their high metal content and close flybys to Earth, objects like 1986 DA and perhaps 2016 ED85 could be possible targets for asteroid mining in the future. In this section we provide an estimate of the amount of metals 
that 1986 DA could contain and how much they could be worth. We have left 2016 ED85 out of this section because of the limited data available for this object.

The analysis presented here should be taken as a rough estimate, since accurate values for the mass, bulk density, and abundance of metal for 1986 DA are unknown. Because of this, several assumptions are made; in 
particular, we will assume that the properties measured from the surface of the asteroid are representative of the whole body. Since this is a relatively small object, this assumption is useful as a first-order approximation. Estimates 
about the cost of developing the necessary technology to extract and deliver the minerals are beyond the scope of this analysis.

We start by calculating the volume of 1986 DA; for this, we will assume that the object is a prolate ellipsoid, which is likely more accurate than assuming a spherical shape. If the amplitude of the lightcurve ($\Delta$m) of the 
asteroid is known, then the ratio of its axes $a/b$ can be determined using the following relationship:

\begin{equation}
\Delta m=2.5log\frac{a}{b}
\end{equation}

For 1986 DA, \cite{2019MPBu...46..304W} found that $\Delta$m = 0.09 mag. If we assume that the radius of the asteroid (1400 m) corresponds to the semimajor axis ($a$), then $b$ = 1288.63 m. The volume of a prolate ellipsoid is 
given by

\begin{equation}
V=\frac{4}{3}\pi ab^{2}
\end{equation}

which yields $V$ = 9.74$\times$10$^{9}$ m$^{3}$. Thus, assuming that the surface bulk density of the asteroid (3790 kg/m$^{3}$) is representative of the whole body, this results in a 
total mass (in metric tons) $M$ = 3.69$\times$10$^{10}$ mt for 1986 DA. From the compositional analysis we determined that this asteroid could have a metal content of $\sim$ 85\%; therefore, the metal mass fraction for this object is 
$M_{metal}$ = 3.14$\times$10$^{10}$ mt.

\begin{table}[h!]
\begin{center}
\caption{\label{t:Table5} {\small Abundance of metals in an average iron meteorite and 6178 (1986 DA). Abundances corresponding to Fe, Ni, Co, and Cu in iron meteorites are from \cite{1975himt.book.....B}. 
The abundances of Au and the PGM are from \cite{1994JGR....9921129K, 1996USGS....2821K}, and correspond to a good 90$^{th}$ percentile in Ir and Pt. Metal values have been calculated only for precious metals, i.e., Au 
and the PGM, and are reported in Trillion USD.}}
\begin{tabular}{cccc}
\tableline

Metal&Abundance&Mass (mt)&Metal value (10$^{12}$ USD) \\  \hline

Fe&90.6 (\% weight)&2.84$\times$10$^{10}$&- \\

Ni&7.9 (\% weight)&2.48$\times$10$^{9}$&-  \\

Co&0.5 (\% weight)&1.57$\times$10$^{8}$&-  \\

Cu&150 (ppm)&4.71$\times$10$^{6}$&-  \\

Ru&20.7 (ppm)&6.49$\times$10$^{5}$&0.1749 \\

Rh&3.9 (ppm)&1.22$\times$10$^{5}$&5.9583 \\

Pd&2.6 (ppm)&8.16$\times$10$^{4}$&1.7035 \\

Os&14.1 (ppm)&4.42$\times$10$^{5}$&0.0161 \\

Ir&14.0 (ppm)&4.39$\times$10$^{5}$&0.5167 \\

Pt&28.8 (ppm)&9.03$\times$10$^{5}$&2.2406 \\

Au&0.6 (ppm)&1.88$\times$10$^{4}$&1.0436 \\

\tableline
\end{tabular}
\end{center}
\end{table}

\begin{figure*}[!ht]
\begin{center}
\includegraphics[height=9cm]{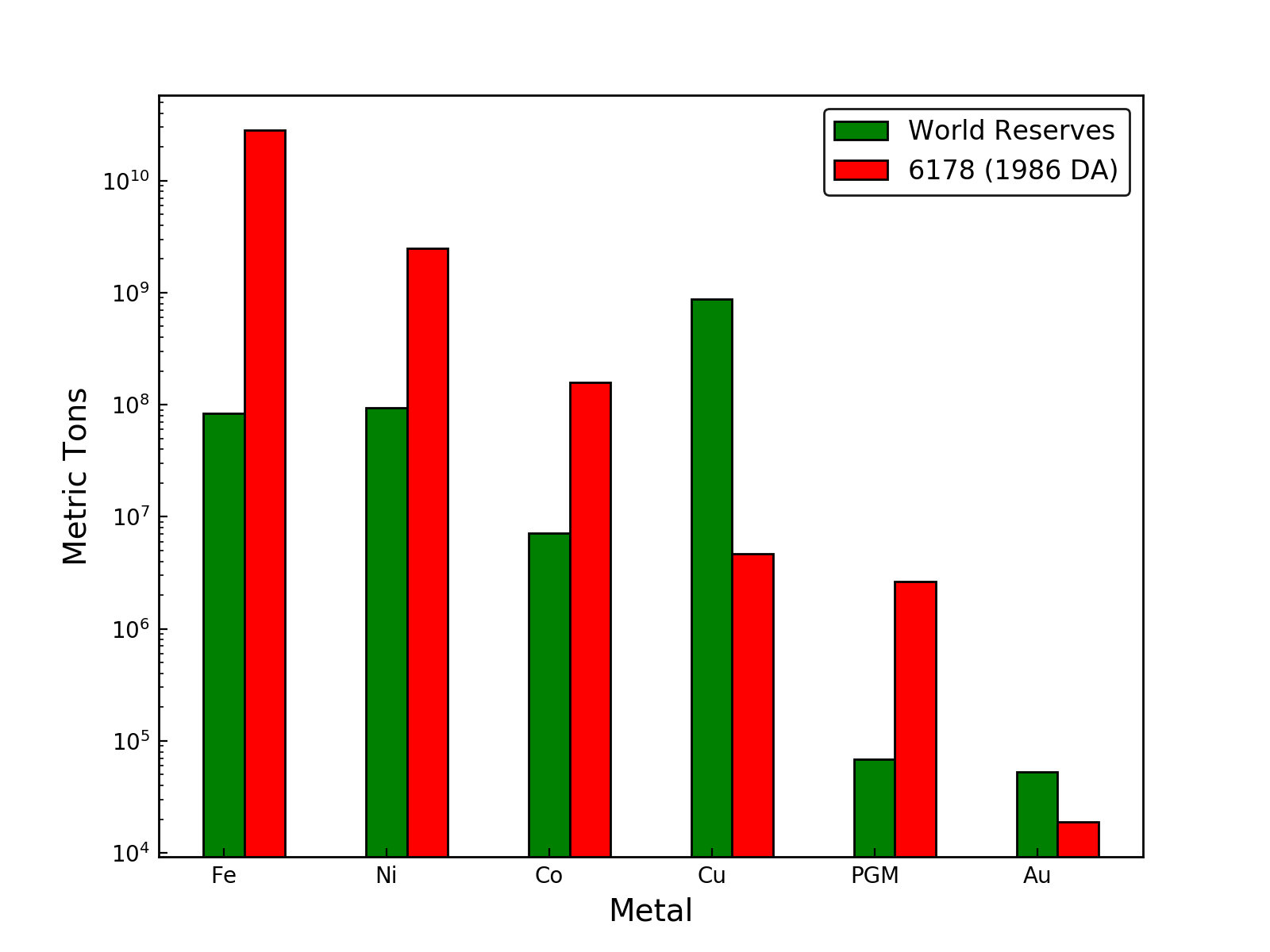}

\caption{\label{f:Figure13} {\small Metal content for 6178 (1986 DA) compared to the reserves worldwide as of 2020. World reserves for each metal were obtained from the U.S. Geological Survey mineral commodity 
summaries 2021.}}

\end{center}
\end{figure*}

Once we have determined the total mass of the metal component, we need to identify the metals that could be present in this object. For this, we use as a reference the composition of the average iron meteorite (Table 5), which 
includes Fe, Ni, and Co as the most abundant metals. We have also included Cu, Au, and the PGM because of their strategic use and value. The mass of each metal was obtained by multiplying its abundance by the total mass of the 
metal component of the asteroid. The amount of each metal for 1986 DA is shown in Table 5. To put these numbers in context, in Figure \ref{f:Figure13} we compare the mass of the metals with the reserves worldwide 
as of 2020. We estimated that the amounts of Cu and Au present in 1986 DA would be lower than those on Earth, whereas the amounts of Fe, Ni, Co, and the PGM would exceed the reserves worldwide.

From this point, determining how much the asteroid is worth would only require multiplying the mass of each metal by its market price. However, increasing the metal supply would have an effect on the prices of these commodities on 
Earth. \cite{1994JGR....9921129K, 1996USGS....2821K} determined that market prices of most of these metals would crash as a result of flooding the market with metals imported from the asteroid. Thus, current market prices need to 
be adjusted to account for this effect. \cite{1994JGR....9921129K} developed an empirical model where the adjusted or deflated prices $p'$ are related to the current prices, $p_{o}$, by

\begin{equation}
p'=p_{o}(P'/P_{o})^{-0.60}
\end{equation}

where $P'$ is the rate of production of the asteroid’s metal plus terrestrial metal and $P_{o}$ is the rate of current production. Using Equation (9), and following \cite{1994JGR....9921129K, 1996USGS....2821K}, we estimated how much 1986 
DA could be worth considering only the precious metals, i.e., Au and the PGM. Assuming that the asteroid is mined and the metals marketed over 50 yr, we found that the annual value of precious metals (in 2021 
U.S dollars) for 1986 DA would be $\sim$\$233 billion. The adjusted prices in Table 5 show that 1986 DA could be worth a total of $\sim$\$11.65 trillion.

\section{Summary}

We carried out a detailed physical and compositional characterization of the NEAs 1986 DA and 2016 ED85. We found that the NIR spectra of both asteroids exhibit 
characteristics distinctive of metal-rich asteroids, i.e., red slopes, convex shapes, and a weak pyroxene absorption band at $\sim$0.93 $\mu$m. Radar observations 
of 1986 DA confirmed that this is a metal-rich asteroid, whereas 2016 ED85 can only be considered as a candidate metal-rich body at this point.

The compositional analysis of 1986 DA and 2016 ED85 showed that these objects have a pyroxene chemistry (Fs$_{40.6\pm3.3}$Wo$_{8.9\pm1.1}$) comparable 
to HEDs and mesosiderites. The intensity of the 0.9 $\mu$m band suggests that both asteroids could be composed of $\sim$15\% pyroxene and 85\% metal.

A comparison between the NIR spectra of the two NEAs and the spectra of M/X-type asteroids in the middle and outer belt showed that the spectral 
characteristics of 1986 DA and 2016 ED85 are consistent with those of the asteroids in the main belt.

We used the NEO model developed by \cite{2017A&A...598A..52G, 2018Icar..312..181G} to determine possible source regions for these bodies. We found 
that the most likely region from which 1986 DA and 2016 ED85 originated is the 5:2 MMR with Jupiter near 2.8 au, with probabilities of 76\% and 49\%, respectively.

We evaluated the scenario in which both NEAs come from the same parent body and resulted from the formation of an asteroid family. This scenario would require the existence of a family with 
similar spectral characteristics, located close to the 5:2 resonance, and capable of delivering bodies with sizes $>$ 3 km to the resonance. We identified the Phaeo family as a possible candidate, although we note that more spectroscopic observations are needed to confirm that it is composed of metal-rich bodies.

The spectra of the NEAs were compared with laboratory spectra of bencubbinite and mesosiderite samples. Differences in the NIR spectra and pyroxene 
chemistry suggest that bencubbinites are not good meteorite analogs for 1986 DA and 2016 ED85. Mesosiderites, on the other hand, were found to have a similar 
pyroxene chemistry and produced a good spectral match when metal was added, suggesting similarities between the parent body of the NEAs and the parent body of these meteorites.

We estimated that the amounts of Fe, Ni, Co, and the PGM present in 1986 DA could exceed the reserves worldwide. Moreover, if 1986 DA is mined 
and the metals marketed over 50 yr, the annual value of precious metals for this object would be $\sim$\$233 billion.



\begin{acknowledgments}

This research work was supported by NASA Near-Earth Object Observations grant NNX17AJ19G (PI: V. Reddy). We thank the IRTF TAC for awarding time to this project, as well as the IRTF TOs and MKSS staff for their support. The 
authors wish to recognize and acknowledge the very significant cultural role and reverence that the summit of Maunakea has always had within the indigenous Hawaiian community. We are most
fortunate to have the opportunity to conduct observations from this mountain. We thank the anonymous reviewers for useful comments that helped improve this paper. 

\end{acknowledgments}

\end{document}